\documentclass[letterpaper,11pt]{article}

\usepackage{amsmath,amssymb}
\usepackage{eepic,epic}
\usepackage{epsfig}
\usepackage{graphicx}
\usepackage{color}
\usepackage{comment}
\usepackage{tabularx}
\usepackage{booktabs}
\usepackage{pifont}
\begin{document}

\title{Regulating Artificial Intelligence\\Proposal for a Global Solution
}
\author{
Olivia J. Erd\'{e}lyi\\
School of Law\\
University of Canterbury\\
Christchurch 8140, New Zealand\\
olivia.erdelyi@canterbury.ac.nz
\and
Judy Goldsmith\\
Department of Computer Science\\
University of Kentucky\\
Lexington, KY 40506, USA \\
goldsmit@cs.uky.edu
}
\date \today

\maketitle

\begin{abstract}
With increasing ubiquity of artificial intelligence (AI) in modern societies, individual countries and the international community are working hard to create an innovation-friendly, yet safe, regulatory environment. Adequate regulation is key to maximize the benefits and minimize the risks stemming from AI technologies. Developing regulatory frameworks is, however, challenging due to AI's global reach and the existence of widespread misconceptions about the notion of regulation. We argue that AI-related challenges cannot be tackled effectively without sincere international coordination supported by robust, consistent domestic and international governance arrangements. Against this backdrop, we propose the establishment of an international AI governance framework organized around a new AI regulatory agency that --- drawing on interdisciplinary expertise --- could help creating uniform standards for the regulation of AI technologies and inform the development of AI policies around the world. We also believe that a fundamental change of mindset on what constitutes regulation is necessary to remove existing barriers that hamper contemporary efforts to develop AI regulatory regimes, and put forward some recommendations on how to achieve this, and what opportunities doing so would present.  
\end{abstract}

\section{Introduction}\label{Intro}

Emerging technologies commonly described by the generic term \emph{artificial intelligence (AI)} are becoming increasingly pervasive in human society. They are extremely transformative, advance rapidly, and affect virtually all aspects of our existence: Self-driving cars are being released on the roads. AI-driven medical diagnosis tools sometimes outperform humans in catching rare diagnoses. Product recommendation systems analyze our needs and optimize our shopping experience. Automated surveillance techniques, killer bots, and other weaponized AI technologies shore up the defenses of our countries. Powerful data mining applications allow us to sift through a wealth of information within a short period of time. AI is revolutionizing financial services with applications reaching from detecting fraud, tax evasion, or money laundering to regulatory technology (RegTech), enhancing regulatory processes like monitoring, reporting, and compliance. The justice system increasingly relies on AI-enabled decision-making systems for predictive policing and sentencing. And the list of examples could go on.

Undoubtedly, some of these technologies can make life a lot easier by providing previously unimaginable benefits. But, given their highly disruptive nature, they also present substantial challenges. Sometimes these problems result from the imperfection of AI applications, as when AI systems produce discriminatory biases, or inappropriate inferences due to biases in training data~\cite{cabitza2017unintended,chouldechova2017fair}. At other times, issues arise because AI is doing its job far too perfectly, as evidenced by the increasing privacy threat posed by pattern recognition applications~\cite{kosinski2017deep}.  Some AI applications lead to human de-skilling \cite{cabitza2017unintended}. Some instantiations of AI are ethically questionable (e.g., child-like sex bots~\cite{strikwerda}) or potentially dangerous (e.g., autonomous weapons systems). Less obviously, AI innovation may also raise broader systemic challenges in the economic~\cite{AGG2019}, legal~\cite{EE2020}, and many other domains, likely forcing us to reevaluate many of our most fundamental ethical, legal, and social paradigms. The overall destabilization of the international community through what is commonly referred to as the \emph{AI race} --- a dangerous competition for technological superiority between AI developers, countries, and regional groupings, encouraging safety and governance corner-cutting and potentially exacerbate existing or even create new conflict situations~\cite{CO2018} --- is yet another danger we face in relation to AI. For a good overview of the contemporary AI landscape and policy environment, see~\cite{RC2017}. 

In order to optimally harness AI's benefits and address its potential risks --- preferably proactively rather than retroactively and in a manner beneficial to all humanity --- it is indispensable to develop adequate policies in relation to AI technologies at the earliest possible stage. Society's growing interest in and anxiety over AI --- fueled by incessant hype and shocking scandals, respectively --- are putting additional pressure on policymakers. The AI community has long been calling for policy action, with criticism getting louder on the growing legal vacuum in virtually every domain affected by technological advancement~\cite{wadhwa2014laws,EM2017,TW2017}. Policymakers around the world have begun to address AI policy challenges. International organizations and groupings have put forward and/or are in the process of developing ethical principles, policy guidelines, and reports concerning AI to provide guidance and assist policymakers' efforts to tackle AI challenges \cite{Asilomar2017,GGAR2018,EC2019,OECD2019,G202019,UNESCO2019,ECAIWP2020}.  Countries have started to launch ambitious strategies to promote the development and commercialization of AI with a view to maintain sustained economic competitiveness after the inevitable global transition to an AI-driven economy~\cite{JN2017,RV2017,OECDPSI,OECDAIS,FoLI2020,ECAIW}. There are also many academic, joint public-private sector, and other venues that support governments in promoting AI R$\&$D.  Examples include the International Association for Artificial Intelligence and Law, the Partnership on AI, The Future Society, the Artificial Intelligence Forum of New Zealand, and SPARC in the EU. More indirectly, cutting-edge tech firms like Amazon, Apple, Baidu, Google, Facebook, IBM, Microsoft, 
and others are also instrumental in shaping AI policies. Courts, faced with first AI-related disputes, contribute to clarifying situations, although some decisions reflect a complete lack of technological expertise~\cite{CJEU:160/15,LGH}.  Such decisions add to, rather than decrease, confusion and underscore the need for interdisciplinary cooperation and improving policymakers' AI expertise~\cite{RC2017}. 

These diverse societal stakeholders enthusiastically engage in shaping our future with AI. Yet the question remains: Do these applications really make human society more efficient, better, or safer? Or is AI rather a looming menace that will ultimately destroy mankind~\cite{SK2015}? In any case, both the AI revolution and the challenges it presents to society are very real and policymakers' efforts to do something about it must continue. 

The importance of \emph{adequately} regulating AI cannot be overstated. History and economic research shows that societal benefits from technological innovation --- including AI --- cannot be taken for granted, but are largely determined by the quality of the market-structuring regulatory environment~\cite{KS2019}: Appropriate regulation corrects market failures by incentivizing socially optimal behavior, ensuring that the benefits of innovation are equally distributed across society. Poorly designed or inappropriately implemented policy measures may, however, impair the status quo by aggravating inequalities and generating tensions between the winners and losers of innovation. In some (perhaps many) cases, some or all segments of society are unnecessarily disadvantaged by uses of AI-driven technology. Inadequate regulatory interventions and protracted periods of uncertainty during regulatory adjustments may also irreversibly destroy society's trust in new technologies. This, in turn, may thwart their societal adoption or even annihilate entire emerging markets, withholding potentially substantial benefits from society~\cite{TP2018}. 

Once an issue --- such as the emergence of AI technologies --- is constructed as a problem and the need for regulation is correctly identified, the next question is how we come up with a regulatory regime that stimulates trust, enables innovation yet also provides safety, and yields socially optimal outcomes? What factors impact on the efficiency of regulation? Addressing them all would not be possible in one paper (but see~\cite{BCL2012} for a comprehensive treatise). Thus, we will concentrate on three interrelated issues that are paramount in the AI context: (1) Why the nature of lawmaking commands international coordination, (2) the importance of institutional architecture in determining the quality of regulation, and (3) practical communication challenges hindering contemporary AI policy development and related regulatory design questions. 

Regarding the nature of lawmaking, a number of peculiarities should be considered so any newly conceived legal norms governing AI become truly authoritative, that is, accepted as legitimate and institutionalized. Otherwise we risk creating merely formal or symbolic rules without any impact on normative orientations and behavior~\cite{H2008}. While regulatory initiatives are predominantly propelled by nation states to address problems at a domestic level, caution is advised with purely national approaches. Whenever the regulation of an issue has externalities that transcend national boundaries --- as does AI regulation --- differing domestic approaches tend to conflict, raising significant difficulties for those affected by more than one regime. An additional problem in such cases is that domestic policymakers' ability and willingness to control the effects of their actions in foreign jurisdictions are limited. These problems are then typically perceived as transnational in scope, with the consequence that actors increasingly deem national rules inapt to provide suitable solutions. This discrepancy between the transnational nature of a problem and the national character of the law governing it creates pressures for transnational regulation. Aligning the scope of regulatory coordination with the reach of externalities has a number of benefits. First, it is the only way to effectively control those externalities. Second, it also facilitates welfare enhancing AI adoption instead of aggravating already pronounced worldwide inequalities, which, in turn, could help increase social and political support~\cite{KS2019,BCL2012}. Transnational legal ordering, however, is characterized by a set of complex, recursive, multi-directional processes, which follow their own logic and crucially affect norms' authority~\cite{HS2015}. Although the legitimacy of legal norms has predominantly social rather than moral roots, their acceptance also depends on the degree to which they conform to prevalent moral values of a given society and --- in the transnational context --- are able to bridge the gaps between conflicting morals of different cultures. Ethical considerations are thus an essential part of any regulatory endeavor.

On the matter of institutional architecture, generally, people immediately associate regulation with its most visible aspect: the actual rules produced by regulators. As the tools of regulation, rules are admittedly important. But such a narrow conception of regulation ignores that developing rules to address a given problem is a small segment of the full regulatory process --- a series of tasks from detecting an anomaly and devising an adequate regulatory response to effective supervision, enforcement, as well as continuous assessment and adaptation of regulatory regimes to ensure optimal performance, all of which must seamlessly complement each other so that rules can be successful~\cite{BCL2012}. Throughout this paper, when we speak of the efficiency or quality of \emph{regulation}, or use the terms \emph{regulatory regime} and \emph{regulatory environment}, we intend to refer to this comprehensive regulatory process. One of those less palpable components of creating an enabling regulatory environment --- for AI as much as any other issue --- is \emph{institutional architecture} (sometimes also referred to as \emph{governance} framework or arrangements), which structures the collaboration of all parties involved in policymaking. In part because the significant workload associated with the urgency to issue tangible pieces of regulation typically exhausts regulators' capacities, architectural design questions are often neglected or even overlooked. This is unfortunate, as they strongly affect the quality of both domestic and international AI regulation~\cite{BCL2012,AS2000}. We argue that it makes a huge difference whether international AI policy coordination occurs in an ad-hoc, voluntary manner, or is streamlined by robust, consistent national and international governance frameworks. 

Turning to communication challenges and related regulatory design questions, another prevalent misconception in relation to regulation is that it is some necessary evil that the state imposes upon society to safeguard order and influence behavior in desired directions. Again, while not inherently wrong, this view overlooks the modern reality of regulation, which is much more a decentered, dynamic process of co-creation than a purely state-driven enterprise~\cite{JB2001}. Due to the immense complexity of most modern regulatory domains, regulators generally lack both expertise and resources to face regulatory challenges by themselves. As a consequence, innovative, hybrid regulatory settings and strategies leveraging the collaboration and expertise of multiple societal stakeholders are typically superior to their traditional counterparts, and increasingly taking over as the default mode of regulation~\cite{BCL2012}. From this follows that regulation presents vital opportunities for diverse societal stakeholders to instill their interests into regulatory processes early on. This improves the odds of creating rules and regulatory frameworks --- which may be binding or not depending on parties' preferences --- that more adequately reflect aggregate collective preferences. We posit that this imperfect understanding of the notion of regulation is currently prevailing among stakeholders. It clouds the aforementioned opportunities from their view and evokes an overly cautious or even hostile attitude in them towards regulation, significantly impeding AI regulatory efforts worldwide. We believe that targeted emphasis on educating parties and clear communication about regulatory intentions, expectations, and opportunities could significantly alleviate these problems. In light of the fact that in AI regulation both expertise and resource problems are heightened due to the field's complexity and rapid development, we urge for using these insights and see regulation as an opportunity rather than an obstacle. 

Against this background, we hold the view that efforts to develop AI policies, should, from the very beginning, be coordinated and supported by adequate national, regional, and international governance frameworks to avoid risks and inefficiencies stemming from the imperfect interaction of fragmented domestic regulatory approaches. To date, such frameworks are missing. AI policies are developed with limited levels of coordination between governments and various academic and industry groupings. The regulatory purviews of agencies involved in policymaking processes are not clearly delineated, and discussions on issues of institutional architecture design are, if at all, in preliminary stages both within governments and across various regional and international fora. In the preliminary version of this paper~\cite{EG2018}, we proposed the establishment of a new intergovernmental organization --- potentially named International Artificial Intelligence Organization (IAIO) --- to serve as an international forum for discussion and engage in standard setting activities. Marchant and Wallach~\cite{MW2018} cultivate a similar idea proposing to set up what they term governance coordination committees (GCCs) either at the national or international level, as appropriate, depending on the issue area addressed. We suggested the IAIO should unite a diverse group of stakeholders from public sector, industry, and academic organizations, whose interdisciplinary expertise can support policymakers in the overwhelming and crucially important task of regulating this novel, immensely complex, and largely uncharted area. Our hope was that such a wide-scale, in-depth cooperation among all interested stakeholders at this early stage would put national and international policymakers in the position to take proactive action instead of lagging behind technological innovation with potentially devastating implications. Yet, establishing a new body is not necessarily the best and surely not the only option to achieve those goals: Since then, a number of new initiatives led by existing and new groupings have emerged that could step up to assume this function or spearhead discussions considering the pros and cons of repurposing other existing bodies or establishing a new agency. The French-Canadian initiative to establish an International Panel on Artificial Intelligence (IPAI) --- renamed Global Partnership on AI, (GPAI) --- the Organisation for Economic Cooperation and Development's (OECD) new AI Policy Observatory (OECD.AI), the Global Governance on AI Roundtable (GGAR) hosted by the World Government Summit (WGS), and a new work stream on AI within the United Nations Educational, Scientific and Cultural Organization (UNESCO) are among the most prominent examples. Thus, in this paper we will also consider how these new developments may tie in with our proposal.

The paper will proceed as follows: Section~\ref{TLO} will present a brief analysis of transnational normmaking processes. This will provide the foundations for our proposal --- introduced in  Section~\ref{IAIO} --- to establish a global AI governance framework organized around either a new or an existing but repurposed intergovernmental organization as the lead global AI policy body. Section~\ref{CRC} will look into the above mentioned communication challenges that currently hamper regulatory efforts in the AI space and discuss related regulatory design issues in a hope to improve the situation. Finally, we summarize our thoughts in a short conclusion.

\section{Dynamics of Transnational Lawmaking}\label{TLO}

In response to economic and cultural globalization, legal, political science, and sociology scholarship have made many
attempts to capture processes of various forms of transnational social ordering. Examples include the traditional, purely state-centric legal notion of \emph{international law} with a dichotomous view towards national and international law; \emph{global law}, which refers to legal norms of universal scope while also acknowledging the role of non-state actors in normmaking; \emph{transnational law}, which can have several connotations in reference to norms with a more than national but less than global purview; the concept \emph{regime theory} developed by international relations scholars, which is likewise state-centric and has a sole focus on international political processes without any regard to the impact of domestic politics or law's normativity; the sociological \emph{world polity theory}, which studies the diffusion of legal norms and assumes that global conceptual models frame national societies in one-dimensional top-down processes; and \emph{transnational or global legal pluralism}, which emphasizes the coexistence of different legal orders and normative contestations among them. Giving a comprehensive overview (including references) of the respective merits and limitations of existing theories, Shaffer~\cite{S2010} and Halliday and Shaffer~\cite{HS2015} introduce a further, socio-legal notion termed \emph{transnational legal order (TLO)}. It builds on these approaches and is defined as a social order of transnational scope consisting of ''a collection of formalized legal norms and associated organizations and actors that authoritatively order the understanding and practice of law across national jurisdictions.'' We explain the determinants influencing transnational legal processes through which legal norms are constructed, conveyed, and institutionalized based on the concept of TLO, owing to its ability to highlight both the legal and institutional aspects justifying the proposed governance framework. 

Disaggregating the above definition into two parts --- (1) formalized legal norms produced solely or partially by some type of transnational legal organization or network, which are (2) aimed at inducing changes within nation-states --- we will first give account of the bewildering variety of governance arrangements characterizing modern international relations, and then illuminate the complex ways in which legal norms interact and institutionalize. The terms \emph{international} and \emph{transnational} shall be used interchangeably, referring to norms and institutions spanning national boundaries (whether global or geographically more restricted in scope). 

Turning to the first element of our TLO definition, both the norms and the institutions issuing them come in a diverse array of forms. Norms are contained in various formal texts of softer or harder legal character such as treaties, conventions, codes, model laws, standards, administrative rules and guidelines, and judicial judgments. International institutionalization displays a similar diversity featuring public intergovernmental (also called international) organizations (IGOs or IOs) and private non-governmental organizations (NGOs) of varying levels of formality~\cite{JK2015,SB2011,VS2013}. The widespread use of both hard and soft legalization in international governance begs the questions of what \emph{hard} and \emph{soft law} are and what drives actors' choices between disparate legal and institutional settings. 

Note that the existing literature is divided on what constitutes hard and soft law. Some authors concentrate on legal rules' binding quality either in binary terms or along a continuum between fully binding legal instruments and purely political, non-binding arrangements, while others focus on their ability to impact behavior.
For a good overview see~\cite{GM2010,SP2010,AS2000}. Because it includes both the legal norms and the institutional arrangements responsible for their development within the scope of its analysis, the most interesting definition for our purposes is the one adopted by Abbott and Snidal. They distinguish hard from soft law along three dimensions, namely (1) the extent of rules' precision, (2) the degree of legal obligation they establish, and (3) whether or not they delegate authority to a third-party decision-maker for interpreting and implementing the law. Hard law refers to legally binding obligations that are either precise or can be made such by adjudication or further clarifying regulation, and that empower a third party to oversee their interpretation and enforcement. Soft law, on the other hand, embodies legal instruments that exhibit some degree of softness along any of these three dimensions. 

Guzman and Meyer~\cite{GM2010} and Abbott and Snidal~\cite{AS2000} provide a very instructive comparison of the relative advantages and disadvantages of hard and soft legalization and the various factors that determine actors' preferences towards different types of international governance arrangements.

Hard legalization is typically characterized by a coherent, established, and formalized institutional and procedural framework to ensure smooth implementation, elaboration, and enforcement of commitments. These arrangements are generally perceived as legitimate, resulting in a concomitant enhanced compliance-pull, and backed up by international law that provides international actors readily available mechanisms (e.g., for recognition or enforcement) to order their relations. The combination of these factors enhances the credibility of commitments by constraining opportunistic behavior and increasing the costs of reneging; reduces post-contracting transaction costs by restricting/constraining attempts to alter the status quo by way of frequent renegotiation, persuasion, or coercive behavior; allows parties to pursue political strategies through legal rather than political channels at low political cost; and solves problems of incomplete contracting by vesting an administrative or judicial institution with power for interpreting and clarifying rules intentionally left imprecise in anticipation of unforeseeable future contingencies. Yet, hard legalization comes at certain costs: it restricts actors' behavioral freedom, entails potentially severe sovereignty implications, and is less effective in accommodating diversity or adapting to changing circumstances by reason of its relative rigidity.

Thus, in many instances, softer forms of legalization, which offer some of hard law's perks yet alleviate its intrinsic disadvantages through their flexible, more or less informal cooperation mechanisms, may better serve parties' purposes. By relaxing the level of formality along one or more of the dimensions precision, obligation, and delegation, soft legalization minimizes initial contracting costs and facilitates speedy conclusion of agreements. Bargaining problems become less pronounced, negotiation and drafting requires less scrutiny, and there is no need for potentially challenging approval and ratification processes. Thanks to soft legal commitments' malleable cooperation frameworks, parties retain more control over the overall design and organization of their cooperation, incur lower sovereignty costs, and have an easily adjustable system at their disposal to deal with change and uncertainty. Soft law also has a way of evening out power asymmetries by securing and perpetuating powerful actors' interests at lower sovereignty costs, while at the same time shielding the weak from their pressure. Furthermore, soft law is the only directly available instrument to non-state actors for ordering their interactions. Due to their conciliatory properties, softer forms of legalization leave actors time to acquire sufficient information and expertise to gradually test and develop solutions to problems, encouraging collective learning processes and ever deeper cooperation between them --- benefits that plentifully compensate soft law's central weakness: diminished compliance pull.

In conclusion, the choice between harder and softer types of legalization involves a context-dependent tradeoff, which actors should carefully consider on a case-by-case basis. Vabulas and Snidal~\cite{VS2013} describe the pros and cons of institutional formality and the tradeoffs actors face when moving along a broad spectrum of intergovernmental organizational formality --- especially between formal and informal intergovernmental institutions (FIGOs and IIGOs) --- in an analogous fashion. 

These three analyses show that, in general, actors opt for hard law/higher institutional formality when they (1) wish to enter into a binding commitment in issue areas subject to a high degree of consensus, because violations are hard to detect, or parties wish to signalize their intention to engage in sincere cooperation; (2) are willing to accept sovereignty costs stemming from delegating decision-making authority to a central body in order to establish stronger collective oversight over issue areas where the probability of violations is high and monitoring and enforcement is important; (3) put more value on collective control of information, for instance, to unveil violations and increase peer pressure to induce universal compliance; (4) aim for lower long-term transaction costs to effectively tackle recurring or clear-cut issues in standard operating procedures; (5) intend to set up a sophisticated centralized administration to provide legitimacy and stability for supporting complex work processes such as the design and elaboration of norms, coordination involving multiple parties, or judiciary and/or enforcement procedures; (6) are faced with the task of managing routine problems, which is more easily done with established administrative and implementing systems. 

Conversely, soft law/lower institutional formality is preferable when actors (1) want to maintain flexibility to deal with uncertainty, distribution problems, diversity, and changing circumstances; (2) prefer to preserve state autonomy and avoid sovereignty intrusions because welfare gains of cooperation outweigh the potential for defection and opportunism so that agreements are self-enforcing once any focal point for discussions has been established, or when external effects elicited by domestic actions are negligible; (3) insist on avoiding formal transparency mechanisms to maintain closer control of information typically among a more homogeneous group; (4) need lower initial contracting costs to speed up negotiations to be able to act fast (e.g., in crisis situations) or because hard law is not available for lack of consensus; (5) find that minimalist administrative functions are sufficient to support their purposes; (6) must manage high uncertainty (e.g., in initial stages of cooperation or in new/complex issue area) and want to allow themselves time for coordination and establishing common ground without making strong commitments.

Sometimes soft law eventually paves the way towards harder forms of legalization and cooperation becomes increasingly formalized, but in many cases soft legalization and institutional informality have their own justification. In practice, both highly institutionalized FIGOs, such as the United Nations (UN) or World Trade Organization (WTO), IIGOs allowing for laxer cooperation, like the Basel Committee on Banking Supervision (BCBS), private NGOs, for instance the International Chamber of Commerce (ICC), and hybrid forms can be fairly successful and instrumental actors in international lawmaking. 

Table~\ref{T1} gives an overview of the above outlined six tradeoffs actors have to weigh when choosing their desired level of legalization/institutional formalization.

\begin{table}[t]
\centering
\begin{tabular}{|p{7cm}|p{7cm}|} 

\hline
Hard Law/High Institutional Formality & Soft Law/Low Institutional Formality \\

\hline
\hline
binding commitment  &  flexible cooperation arrangements \\ 

\hline
delegation/high sovereignty costs & state autonomy/low sovereignty costs   \\    

\hline
collective control of information & close control of information \\

\hline
low long-term transaction costs & low initial contracting costs \\ 

\hline
complex centralized administration & minimalist administrative functions \\

\hline
routine management  & crisis/uncertainty management  \\

\hline
\end{tabular}
\caption{Tradeoffs in legalization/institutional formality.}
\label{T1}
\end{table}

Moving on to the second part of our TLO definition, transnational legal norms directly or indirectly pursue the ultimate goal to induce shifts in countries' policies and individuals' normative preferences through various formal or informal channels. This generates convoluted, recursive cycles of international lawmaking processes across diverse transnational and national fora, until norms eventually settle and institutionalize~\cite{S2010}. Halliday and Shaffer~\cite{HS2015} note that transnational norm-making may encounter difficulties in the following situations: First, actors may find themselves caught up in diagnostic struggles over the framing of problems, which favors particular alliances and antagonisms, supporting diagnoses reflecting the respective interests of these groupings. Second, domestic implementation of transnationally agreed rules is frequently thwarted and a new cycle of lawmaking is triggered by parties who are influential at the national level, but are not represented or are unsuccessful in international negotiations and therefore refuse to accept such norms as legitimate --- a situation referred to as \emph{actor mismatch}. Third, in their endeavor to reach widely accepted compromises, parties often resort to vague language or leave delicate issues unresolved in their agreements. The resulting ambiguity and built-in contradictions of transnational norms open avenues for nationally fragmented, likely conflicting implementation, again calling for further transnational lawmaking to eliminate related problems. 

Inspired by Shaffer~\cite{S2010}, we now describe the recursive processes of international lawmaking, which encompass mutual interactions both vertically between transnational and domestic venues, as well as horizontally among various TLOs. Vertically, transnational norms impact states in a process referred to as \emph{state change}. Their impact can encompass the whole or parts of the state (\emph{location of change}), it may occur in a slow, progressive process or abruptly owing to unexpected circumstances (\emph{timing of change}), and across five interrelated dimensions. The most obvious aspect of state change is the dynamic evolution of domestic legal systems elicited by the formal national enactment of transnational law. Formal enactment may or may not have a substantial effect on rules' practical implementation depending on the extent to which the transnationally induced change is viewed as legitimate. In more subtle ways, however, these primary legal changes set much broader systemic transformations in motion with potentially heavy social repercussions. For one thing, they continuously reshape established governance models by altering the allocation of functions between the state, the market, and other forms of social ordering. At times, this prompts more state intervention, giving birth to new public and public-private hybrid agencies, while at other times it propels deregulatory tendencies resulting in a retreat of state administration and simultaneous engagement in self-regulation by the private sector. Moreover, transnational legal processes are often responsible for revamping states' institutional architecture. They shift power between different branches of government and upset the division of responsibilities among existing state institutions, sometimes giving rise to new additions to the institutional landscape. It is not hard to see that domestic systems may starkly differ, and such fragmentation often entails devastating consequences in issue areas with cross-border effects. These legal, governance, and institutional changes directly affect individuals by reconfiguring markets for professional expertise, which, in turn, feeds back into the adaptation of governance models by, e.g., a move towards more technocratic forms of governance. This highlights an important, yet admittedly somewhat elusive point, namely that not only institutions but also individuals --- acting as conduits facilitating the diffusion of transnational norms --- play a crucial role in domestic and transnational lawmaking. The fifth domain of state change concerns the modifications in patterns of association and mechanisms of accountability across various national and international sites of governance. These shifts ultimately shape individuals' legal culture and consciousness, as well as their expectations towards the state, triggering new processes of state change where these views conflict with the prevailing state of affairs. 

The extent, location, and timing of state change hinges on three clusters of factors pertaining to the TLO's nature, its relation to the receiving state, and the receiving state's peculiarities. First, TLOs are generally better received if perceived legitimate, i.e., norms are adopted by respected actors with preferably similar interests, in a fair (especially non-coercive) process, and they effectively tackle designated target problems. Myriads of international and national, state and non-state actors interact in complementary or conflicting ways in shaping every aspect of transnational lawmaking. They seek to legitimize rules that serve their purposes and delegitimize those running against them. Powerful players typically dictate the outcome of such struggles. TLOs are more likely to have real behavioral impact if they consist of accepted, clear, and well-understood norms. As discussed above, binding hard law does not necessarily score better in this respect. In a large part, TLOs' coherence is a function of the quality and quantity of their horizontal interaction, and can be threatened where significantly overlapping TLOs interact in an antagonistic rather than complementary fashion~\cite{SP2010}.

As far as TLOs' relation to the receiving state is concerned, powerful actors sometimes resort to coercive measures to impose their will on weaker countries. However, because coercion irrevocably destroys norms' legitimacy, changes forced on states in this manner are at best symbolic and short-lived before they are successfully blocked at the stage of domestic implementation. Another essential prerequisite for the sustainability of transnationally triggered change is the support of intermediaries, who link transnational and national lawmaking processes and are deeply familiar with the interests of both sides. Whether government representatives, industry specialists, academics, social movement leaders, or professionals employed with various public or private organizations on the national and international platform, these intermediaries are instrumental in coordinating communication, easing tensions, and conveying norms between the national and transnational levels.  

Finally, the single most important condition for transnational legal norms' national acceptance is their conformity with the target country's existing cultural and institutional settings and pursued reform initiatives. It strongly depends on the receiving country's prevailing power configurations, institutional capacities, path dependencies, and cultural disposition, and tends to decrease as the distance between the transnational and national contexts and interests and/or the extent of state change increases.    

This concludes our analysis of transnational legal ordering, highlighting the main factors instrumental in determining transnational legal norms' efficiency in influencing the behavior of states and their various constituencies. Our aim was to show that the choice of governance arrangements for an issue area in question --- in our case AI --- crucially determines the overall efficiency of the regulatory regime governing it. We have also seen that legal norms and lawmaking processes interact in complex ways between the transnational and national levels. This means we must strive to design AI governance frameworks that are consistent across these levels, ensuring that all actors affected by regulation are appropriately represented at some point in the process, and hence willing to accept any rules the frameworks produce. We now turn to our proposal on how an international AI governance framework could look like.

\section{Proposal for a New International Artificial Intelligence Organization}\label{IAIO}

International institutions are the prevalent vehicles of international cooperation in our interconnected world. When a critical mass of states and/or non-state actors feel that transnational cooperation is necessary to solve a problem that is impossible to tackle by isolated national measures, they establish a new IGO or NGO for that particular purpose. Based on legal and international relations definitions in circulation --- see~\cite{JK2015,SB2011,VS2013} --- we define an IGO as a formal entity (1) established by an international agreement governed by international law; (2) with at least three (sometimes two) members --- typically states but increasingly also IGOs; and (3) having at least one organ with a will distinct from that of its members. FIGOs' organizational purpose is laid down in a binding international agreement such as a treaty or a formal legal act of another IGO, their membership is clearly defined in the founding legal act, and they have a permanent and significant institutionalization in place. By contrast, IIGOs operate based on an explicitly shared, but informal expectation about purpose, their membership is not always clear, as members are explicitly associated but only by non-legal mutual acknowledgment, and they do not possess any significant institutionalization. NGOs differ from IGOs in that they are not created by treaty --- meaning they are governed by national rather than international law --- and their membership is made up of non-state actors.

Given the severity and global nature of AI's anticipated impact on humanity, we expect it to join the long line of issue areas requiring interstate cooperation, raising the question of establishing an IGO at some point in the future. Against this background, we propose the creation of the IAIO as a new IGO, which could initially serve as a focal point of policy debates on AI-related matters and --- given sufficient international support --- acquire increasing role in their regulation over time. We start by determining the degree of desired institutional formalization by examining, in turn, the six above elaborated tradeoffs in relation to AI.

\emph{Binding commitment vs. flexible cooperation arrangements}: As pointed out earlier, AI will fundamentally transform human society worldwide. Since this process of transformation is likely to be inescapable for any single state, states will probably wish to cooperate sincerely. Also, violations will be difficult to detect as keeping pace with technological innovation will require considerable technical expertise and capacities, presumably exceeding especially weaker countries' capabilities and evoking severe power asymmetries. While apart from this latter circumstance, these factors speak for hard legal commitments, it must be kept in mind that AI research and AI-human interactions are relatively young phenomena and their novelty severely restricts our ability to anticipate the spectrum and extent of the impending changes, let alone the dimension of the problems they will raise. Many AI instantiations encroach on our most basic rights, 
pose an existential threat,
or bring up profound ethical and social questions,
not to mention that they will utterly and completely upset our legal system. So, we are looking at heated debates among radically diverse parties over a variety of uncertain issues, which may change in rapid and currently unimaginable ways --- conditions that, based on past experience, do not exactly favor international consensus. Therefore, we need all the flexibility we can get to acquire familiarity with the issues at hand, sort out differences, and establish common ground, before we can contemplate drawing up a more binding framework for cooperation.  

\emph{Delegation/high sovereignty costs vs.~state autonomy/low sovereignty costs}: Weaponized AI technologies and certain data mining practices are clearly relevant for national security. As this is a sensitive issue area involving particularly high sovereignty costs, at least initially, states will show reluctance to give up and delegate decision-making authority to the IAIO. In the long run, however, powerful collective oversight and enforcement mechanisms will probably be indispensable in order to curb incentives for violations and opportunistic behavior, which should otherwise be high in light of the major shifts in international power constellations triggered by changes in countries' competitive positions. Also, domestic AI policies will produce significant externalities, affecting other countries. Based on this analysis, it is hard to escape the conclusion that a highly institutionalized organization with binding legislative, dispute resolution, and enforcement authority would be better suited as new international AI regulator. Nevertheless, the political reality remains that until sufficient clarity is reached on the IAIO's precise purpose, membership, the issues to regulate, and the broad directions to follow, international consensus supporting such a high degree of institutionalization is off the table. 

\emph{Collective control of information vs.~close control of information}: History shows that states are generally cautious about sharing information on fate-changing technologies, which speaks for close control of information with respect to AI. However, if and when we manage to gather consensus for hard legal commitments (e.g., treaty on certain AI applications), we will probably need to be more forthcoming with certain information to ensure compliance with those instruments. This is again a strong argument in favor of starting cooperation on AI regulation in a softer institutional framework and using soft law instruments, although a move towards harder legalization seems to be desirable over time.

\emph{Low long-term transaction costs vs.~low initial contracting costs}: International discussions on AI are just beginning and powerful states will likely have divergent preferences with respect to the regulation of this high-impact field. Compounded with the difficulties discussed in the context of previous tradeoffs, this makes the prospect of reaching a workable international consensus in the short term rather remote. Yet crucially, swift regulatory response is imperative to prevent proliferating unregulated AI applications from causing social harm and to ensure that the opportunity presented by the rise of AI is harvested to humanity's benefit rather than detriment --- an aim best facilitated by lowering initial contracting costs with soft legalization and low institutional formalization. This is not to say that the idea of setting up a more robust governance framework with standard operating procedures should be abandoned. On the contrary, such a step has merit, but only at a later stage, in possession of sufficient expertise and political consensus to better assess the implications of various policy options and formulate informed policy recommendations.

\emph{Complex centralized administration vs.~minimalist administrative functions}: Similar considerations apply as far as the level of administrative sophistication of the IAIO is concerned. In the initial stage of determining the purpose of the organization, its membership, the issues that need to be regulated, and the backbone of its regulatory agenda, less is probably more. Later, with perhaps binding legal instruments governing selected aspects of AI for a wide membership, work will get more complex, requiring stronger oversight, dispute resolution, and enforcement mechanisms as well as more powerful bureaucratic functions to service them.

\emph{Routine management  vs.~crisis/uncertainty management}: In view of AI's novelty, extreme complexity, unforeseeable evolution, and the controversies it is expected to elicit among a very heterogeneous circle of members, we are up against managing an extraordinarily uncertain issue area. Consequently, we need time and soft legalization's flexibility to establish commonly shared ideas, interests, cooperation mechanisms, and solutions, which can then form the basis of more formalized cooperation arrangements in the future. 

In summary, at least initially, the IAIO should start out as an IIGO displaying a relatively low level of institutional formality. It should use soft law instruments, such as non-binding recommendations, guidelines, and standards, to support national policymakers in the conception and design of AI-related regulatory policies. Its interim goal should be to galvanize international cooperation, fostering internationally consistent AI policy approaches by directly engaging governments in this domain as early as possible, before states develop their own, diverging policies, which may be hard to rescind without political damage. Like many other key IGOs, the IAIO should be hosted by a neutral country to provide for a safe environment, limit avenues for political conflict, and build a climate of mutual tolerance and appreciation. Whether the international community wishes to move towards more formalized cooperation at some point in the future remains to be seen. Diverse institutional choices in other areas of international cooperation suggest that many different settings can be successful. Sometimes informality turns out to be the key to an organization's success. This seems to be the case with the Bank of International Settlements especially during its initial years of operation and World War II, or the BCBS and the different G-Groups at present~\cite{BT2006}. Another common trajectory is when initially informal arrangements turn into formal frameworks of cooperation. A case in point is the General Agreement on Tariffs and Trade's (GATT) gradual transformation into the WTO~\cite{AS2000}. Finally, there are examples for remarkably successful, sustained, complementary, and mutually beneficial cooperation between several organizations of varying institutional formality in the same issue area. This sort of relationship is characteristic for the IMF and various G-Groups in financial regulation, or the Australia Group (AG), an IIGO, and the Organization for the Prohibition of Chemical Weapons (OPCW), a FIGO, in the regulation of chemical and biological weapons~\cite{VS2013}.  So, even though the prospect of a formal global AI agency with regulatory and perhaps also conflict resolution powers is rather remote at the moment, higher or even full institutional formality might become the best option one day.

As mentioned in the Introduction, when we first came up with the idea to propose the establishment of a global AI governance framework, there was virtually no discussion --- let alone one involving governments --- addressing institutional architectural design questions. Instead, many newly-formed non-governmental stakeholders were just starting to work on selected high-priority topics in an uncoordinated manner. This led us to suggest the formation of a new intergovernmental organization, which we dubbed IAIO. However, since then work on AI has intensified in a number of existing organizations, and new actors and initiatives have also appeared on the horizon. This presents a new situation: Instead of a void, we now have a rudimentary institutional structure. Our goal should be to build on these foundations and find an institutional configuration that maximizes incentives for cooperation and minimizes competition between these new international AI policy actors. As long as one of them can take on the role we envisage for the IAIO --- an issue we will now examine --- it is counterproductive to add yet another institution to the existing landscape of potential AI regulators.  

In February 2020, capitalizing on previous AI work done by the organization, the OECD has established a new AI Policy Observatory (OECD.AI)~\cite{OECD.AI}. This inclusive platform for public policy on AI has the purpose to facilitate international dialogue and collaboration between a wide range of stakeholders representing governments, domestic and international regulators, the private sector, academia, the technical community, and civil society. It provides multidisciplinary, evidence-based policy analysis in areas most strongly affected by AI. Since its inception in 1961, the OECD has proven to be a successful global standard-setter in multiple public policy areas --- the most recent case in point are the OECD AI Principles adopted in 2019~\cite{OECD2019}, which constitute the first intergovernmental standard on AI. The organization's global reach and multi-stakeholder approach facilitates the gradual development of widely accepted best practices by encouraging open, international dialogue, comparison of each other's policy responses, and mutual learning. As an intergovernmental, yet informal forum for AI policymaking, OECD.AI provides the necessary flexibility that is required in the initial stages of global AI policy coordination. At the same time, it is backed by the OECD's established, formal institutional arrangements, which --- if the international community so desires in the future --- may help it transition into a more formal vehicle of cooperation. Should the grouping have aspirations to become the undisputed focal point of global AI policymaking, it will have to find a way to officially represent the entire international community rather than just OECD countries. Ultimately, OECD.AI's success and the trajectory of its evolution will depend on the global political climate.

According to the decision of its 40th General Conference in November 2019, UNESCO has also embarked on an ambitious two-year project with a view to draft the first global standards on AI ethics~\cite{UNESCOAI,UNESCOAHEG2020}. This mission statement seems narrower than that of OECD.AI at first sight. However, it has to be seen within UNESCO's broader mandate to build peace through international cooperation in education, science, and culture, and contribute to achieving the Sustainable Development Goals adopted by the UN General Assembly in 2015 as part of the organization's 2030 Agenda for Sustainable Development. Moreover, it has to be noted that UNESCO's mode of operation is also characterized by a highly inclusive multi-stakeholder approach, engaging multifaceted, interdisciplinary expertise and powerful parties from around the globe. Again, political considerations will play a central role in determining the direction in which UNESCO's AI work-stream will develop over time. In any case, this is another informal grouping embedded into and supported by a --- this time truly global --- FIGO, which could decide to broaden its mandate and become the sort of intergovernmental AI policy body we propose to put in place. 

Another potentially interesting new initiative is the Global Governance on AI Roundtable (GGAR)~\cite{GGAR}. The first two editions of the Roundtable in 2018 and 2019 have been hosted by the World Government Summit (WGS) held in Dubai. Likewise employing an international, interdisciplinary, and multi-stakeholder approach, this endeavor's primary objective is to assist the UAE State Minister for AI in developing the UAE's AI strategy, which was announced in 2017. However, by moving the venue of discussion away from developed western countries, which traditionally lead global dialogue in virtually all issue areas, GGAR offers a radically novel and different political context. This may resonate better with certain countries and provide an opportunity to bring parties with previously irreconcilable political positions on board. This hope is reflected in GGAR's other key aim: To serve as a neutral forum that coordinates with a wide range of existing stakeholders vested in global AI policymaking, enabling the international community to shape globally accepted and culturally adaptable norms for AI governance. GGAR is a brand new informal grouping, which --- if it awakens the international communities' interest --- could be morphed into a global intergovernmental AI agency and perhaps evolve into a more formal organization without being constrained by existing path dependencies.

In June 2018, France and Canada announced an initiative to establish an International Panel on Artificial Intelligence (IPAI), later renamed Global Partnership on AI (GPAI)~\cite{MIPAI2018,AG2019}. Once taking up work, the GPAI's mission will be to support and guide responsible, human-centric AI adoption, respecting human rights and ensuring inclusion, diversity, innovation, and economic growth. It is also envisaged to rely on interdisciplinary, multi-stakeholder mechanisms to harness leading global expertise, and collaborate with other international AI policy bodies to facilitate AI research, information sharing, and the development of widely accepted, international best practices. At the present juncture, there is little clarity on the GPAI's organizational structure, relationship to other existing AI policy actors, specific mandate, and political reception, so it is hard to predict if it could grow into a universally accepted, intergovernmental AI policymaking institution.

A last organization we would like to mention here is the World Economic Forum (WEF), in particular its Centre for the Fourth Industrial Revolution (4IR) launched in 2017~\cite{WEF}. The WEF is an NGO founded on the principle of public-private cooperation, which also relies on a global network of expertise, partnering with a wide range of stakeholders from both the public and private sectors to contribute to shaping local, regional, and global policy agendas. 4IR is intended to function as a hub for global, multi-stakeholder cooperation that takes the lead in co-designing and piloting innovative policy frameworks and governance protocols related to emerging technologies, including but not limited to AI. The interesting thing about the WEF is that as a non-governmental entity, it is led by the private sector and only involves governments indirectly. Yet it essentially works with the same methods as the other intergovernmental initiatives listed above. What is more, by focusing on well delineated pilot projects implemented jointly with selected government partners, it has substantial real impact on AI policies and industry practices at the national, regional, and global levels. While as an NGO, the 4IR cannot directly assume the role of a global intergovernmental AI regulatory body, it is definitely an actor with major influence that needs to be taken into account in the design of the suggested global AI governance framework.

There are also several other intergovernmental bodies that do valuable work on AI and are instrumental in steering global AI policies. Examples include the European Commission~\cite{ECworkAI,ECAIW}, the Council of Europe~\cite{CAHAI}, and the G7 and G20 groups. However, as European institutions, the first two represent the EU rather than the international community. As for the informal G-Groups, these have a broader focus than just AI and there are good reasons to preserve them in their current roles as quickly mobilizable, flexible vehicles that devise informal solutions supporting the work of FIGOs in various policy domains. Hence, neither of them constitute adequate fora to serve the purpose we envisage for the IAIO. That said, these are also highly influential stakeholders that need to be involved in any international AI governance arrangements. 

To sum up, OECD.AI, UNESCO's AI group, GGAR, and the GPAI are potentially viable vehicles to take on the role of an intergovernmental AI regulatory agency as an alternative to setting up an entirely new organization in the form of the IAIO. This ends our excursus in the domain of international lawmaking, which also shows that beyond the optimal level of legalization and institutional formality, the proposed new intergovernmental AI policy body --- be it one of the above introduced new players or a newly established IAIO --- must fulfill a number of more subtle requirements to be perceived as a \emph{fair and legitimate} regulator. While leaving the elaboration of details to political decisions and future research, we would like to stress three points: (1) It is necessary to put an IGO in charge of leading international AI regulatory efforts to ensure sufficient government involvement and impact on domestic AI policies, as well as to keep the option open to move towards more formal institutional arrangements. (2) We must maintain an inclusive, interdisciplinary, and multi-stakeholder approach in all aspects global AI policy design, including the initial deliberations related to the IAIO's or its equivalent's establishment, modus operandi, and regulatory agenda. This is the only way to ensure the availability of expertise necessary to effectively tackle AI-related challenges. (3) Finally, it is paramount to choose an institutional setting that adequately reflects all relevant AI actors' interests and existing power constellations to guarantee the widespread acceptance and legitimacy of the proposed global AI governance framework. There is already some collaboration between the policy bodies and other stakeholders introduced in this paper. However, the status quo is much less effective compared to what would be possible if a well-designed and universally accepted global AI governance framework organized around a single agency was introduced. Such a framework could streamline all these stakeholders' efforts ensuring an appropriate division of labor by optimally exploiting synergies in a clear and transparent manner. We are not proposing a complete overhaul of the current system, merely to eliminate inefficiencies by introducing an institutional architecture that maximizes collaboration and minimizes the duplication of tasks and competition between all actors involved. As mentioned earlier, the quality of a governance framework crucially determines the efficiency of the rules it produces: We cannot have efficient and widely accepted transnational AI norms if the adequacy and legitimacy of the governance framework producing them is put into question. Hence, in an ideal world, figuring out the right governance arrangements logically precedes the creation of any rules.

\section{Communication: An Insidious Regulatory Challenge}\label{CRC}

Like all emerging technologies, AI's successful societal adoption hinges on trust, which, in turn, flows from an agile, transparent, and sustainable regulatory environment. As we have seen, governance arrangements are just a part of that, and this paper has only provided some preliminary thoughts on the international setting. Ideally, the proposed global AI governance framework needs to be complemented by robust national AI regulatory regimes, which duly represent national interests and reflect the domestic stakeholder landscape. We respect that --- being shaped by each country's unique political situation as well as cultural and other path dependencies --- national regimes will inevitably display differences. Still, some international coordination in setting up domestic regimes is desirable to prevent discrepancies which may lead to clashes between countries. Leaving these questions aside for now, we would like to direct attention to two instances of what we believe to be ultimately communication challenges: (1) issues around the general perception of the notion of regulation, and (2) interdisciplinary and inter-stakeholder communication barriers. Both are less obvious but highly pernicious problems, which currently heavily stifle efforts to establish AI regulatory regimes at all levels. Thus, both the international community and individual countries may find these insights helpful when designing the fundamental elements of their respective regulatory regimes. 

As noted in the Introduction, people usually think of regulation as binding rules forced on society by the state either in the form of purely domestic regulatory measures or implementing international legal commitments. Especially businesses see regulation as an obstacle designed by the state --- the \emph{enemy} --- to restrict their activities, which they consequently somehow have to work around. Admittedly, there is some truth in this view, as regulation ideally aims to incentivize socially rather than individually optimal behavior, and hence unavoidably restrains activities that threaten to decrease society's welfare. However --- apart from the aforementioned fact that creating rules is just a small part of regulation --- this conception raises at least two additional problems by assuming an adversarial relationship between the state, businesses, and civil society. 

First, it does not necessarily reflect the genuine preferences of these three stakeholder groups --- which may or may not contradict depending on the particular scenario in question. Take AI innovation for instance. At first sight, innovators' interests misalign with those of the state and civil society when it comes to safety considerations. The former may see an opportunity for cost saving by engaging in corner-cuttings, while the latter definitely value safety very highly. Yet this only holds true on the short run, if at all, as safety issues destroy society's trust and hence markets in AI technologies. The resulting situation is against everybody's interests and potentially even welfare decreasing: Innovators are no longer able to derive any financial gains from developing AI, while state and society are deprived from any benefits of innovation. 

Second, it is not in line with modern regulatory reality~\cite{BCL2012,JB2001}. Regulation is no longer a responsibility reserved solely to the state but a decentered process of co-creation, involving multiple stakeholders. The main driver behind this regulatory paradigm shift is the increasing complexity and rapid pace of change of modern regulatory domains, which makes them a prohibitively big challenge for the state --- or any other stakeholder for that matter --- to tackle alone. The recognition that no single party has the knowledge, power, or capacity to effectively control and regulate all segments of society has led modern regulatory theory to move away from the state's regulatory monopoly and advocate a shared responsibility of different actors for regulation instead. According to current regulatory best practices, regulation should be a series of convoluted multi-stakeholder interactions, in which autonomous social actors and government stakeholders are mutually interdependent co-producers of regulation, jointly constructing knowledge and exercising power. The core aim is to create dynamically adaptive patterns of interaction between a multitude of regulatory actors and strategies that best serve the public interest. In widespread opinion, hybrid regulatory mixes and networks deliver the best results in today's globalized and interconnected world. Yet, finding the right blend of institutions and instruments remains challenging. This fundamentally different understanding of regulation dictates the use of diverse regulatory strategies --- incorporating both more state-driven and self-regulatory elements --- and coordination between multiple regulatory actors drawing on wide interdisciplinary and multi-stakeholder expertise. So yes, regulation will always remain state-driven to a certain extent. But it is also an indirect, flexible, and sensitive process of steering, coordinating, balancing, and influencing that not only gives affected societal stakeholders an opportunity to stand up for their interests but cannot be efficiently done without their participation. 

Unfortunately, all this is easier in theory than in practice. It is hard enough to build a regulatory regime that is sufficiently inclusive to ensure that all stakeholders' interests are duly taken into account, really putting public interest first. The trickiest part, however, is to maintain high levels of engagement, satisfaction, and performance in the face of changing conditions. A potential danger stakeholders currently active in AI policymaking may encounter is loss of drive. In the past few years, a large number of public and private bodies around the world have engaged individuals of diverse background into developing solutions to various AI-related challenges. However, lacking clear objectives and adequate coordination within and across such groupings, very little of that energy has been actually transformed into concrete, implementable actions. Low productivity levels have already led to a tangible decline of enthusiasm among participants in some venues, and we expect this trend to continue, especially as the AI-hype continues to wane. This is undesirable, seeing as governments depend on these stakeholders both in terms of expertise and regulatory capacity, and it is also in the latter's best interest to work with governments to implant their preferences into policy initiatives. Hence, there is a sense of urgency in developing regulatory regimes --- perhaps also relying on self-regulatory organizations or otherwise incorporating self-regulatory features --- that set well-defined regulatory objectives and more efficiently coordinate all stakeholders involved. 

To link back to the point of communication challenges raised earlier: In our experience, the true nature of regulation --- including the above explained relationship of co-dependence between different regulatory actors and the benefits that stem from participating in regulatory processes --- are generally not well understood or met by strong skepticism by stakeholders involved in AI policymaking across various fora. Our recommendations to alleviate these problems are twofold: (1) Acknowledging those national and transnational AI policy actors that already comply with modern regulatory best practices, we urge those not yet on this path to embrace and apply these insights in practice when designing their respective AI regulatory regimes. (2) Governments and international policymaking bodies should approach actors they wish to involve in regulatory and policymaking processes with a clear and realistically implementable agenda.  They should unambiguously communicate --- better yet, educate --- them about the nature of regulation, current regulatory best practices, and their intention to follow them. We need to elicit a change of mindset about regulation --- not state-imposed restrictions but an opportunity to co-create regimes and rules that serve aggregate collective preferences --- and give stakeholders reason to trust that these new expectations will be fulfilled in practice.

This leads us to the last point we would like to tackle in this paper, namely interdisciplinary and inter-stakeholder communication barriers. As explained above, all modern regulatory domains have their fair share of complexity. This holds all the more true for AI regulation, as we are dealing with a diverse set of unusually fast-developing technologies, which penetrate virtually all domains of human existence with far-reaching consequences. To make matters worse, due to their complexity, the workings of AI technologies are very hardly accessible for individuals without some technical background. Unfortunately, the vast majority of people --- including numerous policymakers and other stakeholders involved in regulatory and policymaking processes --- grapple with this problem. Owing to the technological intricacies and AI's widespread societal impacts on a global scale, developing sound regulatory approaches requires deep understanding of a wide spectrum of multidisciplinary concepts and internationally concerted, collaborative efforts between multiple stakeholders: policymakers, the AI and other affected industries, academic institutions, and civil society. 

As previously noted, many AI policy actors are aware of these problems and aim to gather the right bundle of expertise to the table. So at least on paper, we seem to be doing just fine. The problem is that the interests, ways of thinking, and modus operandi of different disciplines and stakeholder groups starkly deviate, inducing massive coordination and communication challenges, not to mention frustration. An illustrative example for such tensions are circles of frustration between government, industry, and academic stakeholders. As the drivers of innovation and economic growth, businesses come up with cutting-edge solutions and products to harness AI's benefits. They are much more flexible and faster than academia or the public sector, but do not necessarily have public interest at their heart. Some also lead the way in R$\&$D to underpin their business activity, but many operate on less sound theoretic foundations. As a consequence, businesses tend to be annoyed by governments' lack of expertise, and the slow pace with which both governments and academia operate. Academic stakeholders excel in research, heavily contributing to developing the theoretical foundations that allow for introducing new technologies into society. Often, however, businesses and governments are not aware of existing research results that would solve problems they wrestle with --- a fact of life that understandably upsets academics. Governments, in turn, ideally serve public interest, lack funding and resources comparable to the private sector, and are often also short on expertise. Hence, they are frequently overwhelmed when it comes to assessing the risks and benefits of new technologies, deciding the fate of those technologies, and keeping pace with industry. 

As regards interdisciplinary communication, knowledge transfer between disciplines is far beneath the desired levels due to people's inability to find a way to explain and understand each others' problems and needs. This is counterproductive, as their pieces of knowledge are complementary and cumulatively necessary to successfully tackle AI-related challenges. In our view, a currently severe problem is that many research contributions --- e.g., dealing with various societal impacts of AI --- and policy decisions are made without due consultation of technical experts, even though the researchers or decision-makers obviously lack the necessary technical background to make meaningful contributions to the intersection of AI and their respective fields or to make informed policy decisions. This not only upsets technically literate individuals, but also results in incorrect and technically infeasible scientific recommendations and polices, both of which are very problematic. The former provide flawed foundations for further research and policy action, and also confuse technically illiterate readers. The latter promote the mis-assessment of AI, providing wrong behavioral incentives and tricking society into believing themselves to be protected from potential negative effects of these technologies.  

These problems adversely affect the levels of engagement in regulatory processes and in societal dialogues on AI, society's trust in regulator's ability to design and maintain adequate AI regulatory regimes, and ultimately the acceptance and legitimacy of emerging regimes and norms governing AI. We need to get better at collaborating with and actually listening to each other, and make more responsible judgments about the limits of our own expertise if we are serious about developing adequate AI regulatory regimes and technically feasible rules and policy solutions.

\section{Conclusion}\label{Conclusion}
Given the intensifying worldwide activism in AI regulation and AI's substantial and global impact on human society, we have highlighted some key regulatory considerations and problems to assist domestic and international AI policymakers. We have also proposed a consistent international regulatory framework --- with either a new or a repurposed existing IGO as its focal point --- to streamline and coordinate national policymaking efforts. Learning from past experience in other regulatory fields, our objective is to offer a viable framework for international regulatory cooperation in the issue area of AI to avoid the development of nationally fragmented AI policies, which may lead to international tensions. Should our proposal find sufficient support in the international community, more concrete steps towards setting up the here advocated regulatory framework, and regulatory policies on specific AI issues can be elaborated.

{\small
 \bibliographystyle{acm}
\bibliography{AIRegs}

\begin{thebibliography}{10}

\bibitem{ECworkAI}
{European Commission Work on AI}.
\newblock
  https://ec.europa.eu/digital-single-market/en/artificial-intelligence.

\bibitem{WEF}
{World Economic Forum, Centre for the Fourth Industrial Revolution}.
\newblock
  https://www.weforum.org/centre-for-the-fourth-industrial-revolution/home.

\bibitem{MIPAI2018}
{Mandate for the International Panel on Artificial Intelligence}.
\newblock
  https://pm.gc.ca/en/news/backgrounders/2018/12/06/mandate-international-panel-artificial-intelligence,
  2018.

\bibitem{AS2000}
{\sc Abbott, K.~W., and Snidal, D.}
\newblock Hard and soft law in international governance.
\newblock {\em International Organization 54}, 3 (2000), 421--456.

\bibitem{AGG2019}
{\sc Agrawal, A., Gans, J., and Goldfarb, A.}
\newblock {\em The Economics of Artificial Intelligence: An Agenda}.
\newblock University of Chicago Press, 2019.

\bibitem{Asilomar2017}
{\sc {Asilomar Conference}}.
\newblock Asilomar ai principles.
\newblock https://futureoflife.org/ai-principles/, April 2020.

\bibitem{BCL2012}
{\sc Baldwin, R., Cave, M., and Lodge, M.}
\newblock {\em Understanding Regulation Understanding Regulation: Theory,
  Strategy, and Practice}.
\newblock Oxford University Press, 2012.

\bibitem{JB2001}
{\sc Black, J.}
\newblock {Decentring Regulation: Understanding the Role of Regulation and
  Self-Regulation in a `Post-Regulatory' World}.
\newblock {\em Current Legal Problems 54}, 1 (2001), 103--146.

\bibitem{BT2006}
{\sc Borio, C., and Toniolo, G.}
\newblock One hundred and thirty years of central bank cooperation: a bis
  perspective, 2006.
\newblock BIS Working Papers, No. 197.

\bibitem{EM2017}
{\sc Breland, A.}
\newblock Elon {M}usk: We need to regulate ai before 'it's too late', 2017.
\newblock The Hill, July 17, 2017,
  http://thehill.com/policy/technology/342345-elon-musk-we-need-to-regulate-ai-before-its-too-late.

\bibitem{cabitza2017unintended}
{\sc Cabitza, F., Rasoini, R., and Gensini, G.~F.}
\newblock Unintended consequences of machine learning in medicine.
\newblock {\em {JAMA} 318}, 6 (2017), 517--518.

\bibitem{RC2017}
{\sc Calo, R.}
\newblock Artificial intelligence policy: A primer and roadmap, 2017.
\newblock August 8, 2017, Available at SSRN: https://ssrn.com/abstract=3015350.

\bibitem{CO2018}
{\sc Cave, S., and \'{O}h\'{E}igeartaigh, S.~S.}
\newblock {An AI Race for Strategic Advantage: Rhetoric and Risks}.
\newblock In {\em Proceedings of the 2018 AAAI/ACM Conference on Artificial
  Intelligence, Ethics, and Society\/} (2018), {AAAI/ACM}.
\newblock {Forthcoming. Available at:
  http://www.aies-conference.com/wp-content/papers/main/AIES\_2018\_paper\_163.pdf}.

\bibitem{chouldechova2017fair}
{\sc Chouldechova, A.}
\newblock Fair prediction with disparate impact: A study of bias in recidivism
  prediction instruments.
\newblock {\em Big Data 5}, 2 (2017), 153--163.

\bibitem{CAHAI}
{\sc {Council of Europe}}.
\newblock {Ad-Hoc Committee on Artificial Intelligence (CAHAI)}.
\newblock https://www.coe.int/en/web/artificial-intelligence/cahai.

\bibitem{CJEU:160/15}
{\sc {Court of Justice of the European Union}}.
\newblock Gs media v sanoma media netherlands and others, case c-160/15.
\newblock Court of Justice of the European Union, Judgment of 8 September 2016,
  GS Media v Sanoma Media Netherlands and Others, Case C-160/15,
  ECLI:EU:C:2016:644.

\bibitem{EE2020}
{\sc Erd\'{e}lyi, O.~J., and Erd\'{e}lyi, G.}
\newblock {The {AI} Liability Puzzle and a Fund-Based Work-Around}.
\newblock In {\em Proceedings of the Third AAAI/ACM Conference on Artificial
  Intelligence, Ethics, and Society\/} (2020), pp.~50--56.

\bibitem{EG2018}
{\sc Erd\'{e}lyi, O.~J., and Goldsmith, J.}
\newblock {Regulating Artificial Intelligence: Proposal for a Global Solution}.
\newblock In {\em Proceedings of the First AAAI/ACM Conference on Artificial
  Intelligence, Ethics, and Society\/} (2018), pp.~95--101.

\bibitem{ECAIW}
{\sc {European Commission}}.
\newblock {AI Watch}.
\newblock {https://ec.europa.eu/knowledge4policy/ai-watch\_en}.

\bibitem{EC2019}
{\sc {European Commission}}.
\newblock {Ethics Guidelines for Trustworthy AI}, 2019.
\newblock 8 April 2019.

\bibitem{ECAIWP2020}
{\sc {European Commission}}.
\newblock {White Paper on Artificial Intelligence - A European Approach to
  Excellence and Trust}.
\newblock
  {https://ec.europa.eu/info/sites/info/files/commission-white-paper-artificial-intelligence-feb2020\_en.pdf},
  February 2020.

\bibitem{FoLI2020}
{\sc {Future of Life Institute}}.
\newblock {National and International AI Strategies}.
\newblock https://futureoflife.org/national-international-ai-strategies/, April
  2020.

\bibitem{G202019}
{\sc G20}.
\newblock {G20 Ministerial Statement on Trade and Digital Economy}, 2019.
\newblock Meeting of 8 and 9 June 2019.

\bibitem{AG2019}
{\sc Gurr\'{i}a, A.}
\newblock {Remarks for G7 Leaders Summit: Digital Economy and Artificial
  Intelligence}.
\newblock
  https://www.oecd.org/about/secretary-general/artificial-intelligence-g7-summit-france-august-2019.htm,
  August 2019.

\bibitem{GM2010}
{\sc Guzman, A.~T., and Meyer, T.~L.}
\newblock International soft law.
\newblock {\em Journal of Legal Analysis 2}, 1 (2010), 171--225.

\bibitem{HS2015}
{\sc Halliday, T., and Shaffer, G.~C.}
\newblock Transnational legal orders.
\newblock In {\em Transnational Legal Orders}, T.~Halliday and G.~C. Shaffer,
  Eds. Cambridge University Press, 2015, pp.~3--72.

\bibitem{H2008}
{\sc Hurd, I.}
\newblock Theories and tests of international authority.
\newblock In {\em The UN Security Council and the Politics of International
  Authority}, B.~Cronin and I.~Hurd, Eds. Routledge, 2008, pp.~23--39.

\bibitem{JK2015}
{\sc Klabbers, J.}
\newblock {\em An Introduction to International Organizations Law}.
\newblock Cambridge University Press, 2015.

\bibitem{SK2015}
{\sc Kohli, S.}
\newblock Bill {G}ates joins {E}lon {M}usk and {S}tephen {H}awking in saying
  artificial intelligence is scary, 2015.
\newblock QUARTZ, January 29, 2015,
  https://qz.com/335768/bill-gates-joins-elon-musk-and-stephen-hawking-in-saying-artificial-intelligence-is-scary/.

\bibitem{KS2019}
{\sc Korinek, A., and Stiglitz, J.~E.}
\newblock {Artificial Intelligence and Its Implications for Income Distribution
  and Unemployment}.
\newblock In {\em The Economics of Artificial Intelligence: An Agenda}, A.~K.
  Agrawal, J.~Gans, and A.~Goldfarb, Eds. University of Chicago Press, 2019.

\bibitem{kosinski2017deep}
{\sc Kosinski, M., and Wang, Y.}
\newblock Deep neural networks are more accurate than humans at detecting
  sexual orientation from facial images., 2017.
\newblock PsyArXiv.

\bibitem{LGH}
{\sc {Landgericht Hamburg}}.
\newblock Az. 310 o 402/16.

\bibitem{JN2017}
{\sc New, J.}
\newblock How governments are preparing for artificial intelligence, 2017.
\newblock Center for Data Innovation, August 18, 2017,
  https://www.datainnovation.org/2017/08/how-governments-are-preparing-for-artificial-intelligence/.

\bibitem{OECD2019}
{\sc {Organisation for Economic Co-operation and Development (OECD)}}.
\newblock {Recommendation of the Council on Artificial Intelligence}, 2019.
\newblock Adopted on 22 May 2019, C/MIN(2019)3/FINAL.

\bibitem{OECD.AI}
{\sc {Organisation for Economic Co-operation and Development (OECD) AI Policy
  Observatory}}.
\newblock {https://oecd.ai}.

\bibitem{OECDAIS}
{\sc {Organisation for Economic Co-operation and Development (OECD) AI Policy
  Observatory}}.
\newblock {Countries and Initiatives Overview}, 2020.

\bibitem{OECDPSI}
{\sc {Organisation for Economic Co-operation and Development (OECD) Observatory
  on Public Sector Innovation}}.
\newblock {AI Strategies and Public Sector Components}.
\newblock https://oecd-opsi.org/projects/ai/strategies/.

\bibitem{TP2018}
{\sc Pearl, T.}
\newblock {Compensation at the Crossroads: Autonomous Vehicles and Alternative
  Victim Compensation Schemes}.
\newblock In {\em Proceedings of the 29th European Regional Conference of the
  International Telecommunications Society (ITS): ''Towards a digital future:
  Turning technology into markets?''\/} (2018).

\bibitem{SB2011}
{\sc Schermers, H.~G., and Blokker, N.~M.}
\newblock {\em International Institutional Law: Unity Within Diversity}.
\newblock Martinus Nijhoff Publishers, 2011.

\bibitem{S2010}
{\sc Shaffer, G.~C.}
\newblock Transnational legal process and state change: Opportunities and
  constraints.
\newblock {\em Minnesota Legal Studies Research Paper}, 10-28 (2010), 1--53.

\bibitem{SP2010}
{\sc Shaffer, G.~C., and Pollack, M.~A.}
\newblock Hard vs. soft law: Alternatives, complements and antagonists in
  international governance.
\newblock {\em Minnesota Law Review 94}, 3 (2010), 706--799.

\bibitem{strikwerda}
{\sc Strikwerda, L.}
\newblock Legal and moral implications of child sex robots.
\newblock In {\em Robot Sex}, J.~Danaher and N.~Mc{A}rthur, Eds. MIT Press,
  2017.

\bibitem{UNESCOAI}
{\sc {United Nations Educational, Scientific and Cultural Organization
  (UNESCO)}}.
\newblock {https://en.unesco.org/artificial-intelligence}.

\bibitem{UNESCOAHEG2020}
{\sc {United Nations Educational, Scientific and Cultural Organization (UNESCO)
  Ad-Hoc Expert Group (AHEG)}}.
\newblock {Toward a Draft Text of a Recommendation on the Ethics of Artificial
  Intelligence}.
\newblock https://unesdoc.unesco.org/ark:/48223/pf0000373199, 10 April 2020.

\bibitem{UNESCO2019}
{\sc {United Nations Educational, Scientific and Cultural Organization (UNESCO)
  World Commission on the Ethics of Scientific Knowledge and Technology}}.
\newblock {Preliminary Study on the Ethics of Artificial Intelligence}.
\newblock https://en.unesco.org/artificial-intelligence/ethics, April 2020.

\bibitem{VS2013}
{\sc Vabulas, F., and Snidal, D.}
\newblock Organization without delegation: Informal intergovernmental
  organizations (iigos) and the spectrum of intergovernmental arrangements.
\newblock {\em The Review of International Organizations 8}, 2 (2013),
  193--220.

\bibitem{RV2017}
{\sc Viola, R.}
\newblock The future of robotics and artificial intelligence in europe, 2017.
\newblock European Commission, Digital Single Market Blog Post, February 16,
  2017,
  https://ec.europa.eu/digital-single-market/en/blog/future-robotics-and-artificial-intelligence-europe.

\bibitem{wadhwa2014laws}
{\sc Wadhwa, V.}
\newblock Laws and ethics can't keep pace with technology.
\newblock {\em Massachusetts Institute of Technology: Technology Review 15\/}
  (2014).

\bibitem{MW2018}
{\sc Wallach, W., and Marchant, G.}
\newblock {An Agile Ethical/Legal Model for the International and National
  Governance of AI and Robotics}.
\newblock In {\em Proceedings of the 2018 AAAI/ACM Conference on Artificial
  Intelligence, Ethics, and Society\/} (2018), {AAAI/ACM}.
\newblock Forthcoming. Available at:
  http://www.aies-conference.com/wp-content/papers/main/AIES\_2018\_paper\_77.pdf.

\bibitem{TW2017}
{\sc Walsh, T.}
\newblock {EU} parliament: Consultation on robotics and artificial intelligence
  - summary and contributions, 2017.
\newblock
  http://thefutureofai.blogspot.de/2017/10/eu-parliament-consultation-on-robotics.html.

\bibitem{GGAR}
{\sc {World Government Summit}}.
\newblock {Global Governance on AI Roundtable (GGAR)}.
\newblock https://ggar.worldgovernmentsummit.org/en.

\bibitem{GGAR2018}
{\sc {World Government Summit}}.
\newblock {Summary Report 2018 Global Governance of AI Roundtable}.
\newblock
  https://www.worldgovernmentsummit.org/api/publications/document?id=ff6c88c5-e97c-6578-b2f8-ff0000a7ddb6,
  2018.

\end{thebibliography}
} 

\end{document}